\documentclass[hyper]{JHEP3}
\usepackage{amsfonts}
\usepackage{amscd}
\usepackage{amssymb}
\usepackage{amsmath}
\usepackage{epsfig,cite}
\usepackage{latexsym}

\def \be  {\begin{equation}}
\def \ee  {\end{equation}}
\def \ba  {\begin{eqnarray}}
\def \ea  {\end{eqnarray}}
\def \bb  {\begin{thebibliography}}
\def \eb  {\end{thebibliography}}
\def \lab #1 {\label{#1}}
\newcommand\PT{\mathbb{PT}}
\newcommand\cA{\mathcal{A}}

\newcommand\cL{\mathcal{L}}

\newcommand\cO{\mathcal{O}}

\newcommand\cN{\mathcal{N}}
\newcommand\C {\mathbb{C }}

\newcommand\CP {\mathbb{CP}}
\newcommand\RP {\mathbb{RP}}

\newcommand\rd{\mathrm{d}}

\newcommand\e{\mathrm{e}}
\newcommand\im{\mathrm{i}}
\newcommand\la{\langle}
\newcommand\ra{\rangle}
\newcommand\del{\partial}
\newcommand\delbar{\bar{\partial}}
\newcommand\MHVbar{\overline{\mbox{MHV}}}

\title{A Direct Proof of BCFW Recursion for Twistor-Strings}
\author{David Skinner\\
	{\it Perimeter Institute for Theoretical Physics,\\ 
	31 Caroline St., Waterloo, ON, N2L 2Y5,\\
	 Canada}}

\abstract{
This paper gives a direct proof that the leading trace part of the genus zero twistor-string path integral obeys the BCFW recursion relation. This is the first complete proof that the twistor-string correctly computes all tree amplitudes in maximally supersymmetric Yang-Mills theory. The recursion has a beautiful geometric interpretation in twistor space that closely reflects the structure of BCFW recursion in momentum space, both on the one hand as a relation purely among tree amplitudes with shifted external momenta, and on the other as a relation between tree amplitudes and leading singularities of higher loop amplitudes. The proof works purely at the level of the string path integral and is intimately related to the recursive structure of boundary divisors in the moduli space of stable maps to $\mathbb{CP}^3$.}

\begin{document}

%%%%%%%%%%%%%%%%%%%%%%%%%%%%%%%%%%%%%%%%%%%%%%%%%%%
%%%%%%%%%%%%%%%%%%%%%%%%%%%%%%%%%%%%%%%%%%%%%%%%%%%

\section{Introduction}
\label{sec:intro}

There is by now little doubt that the twistor-string theory captures the complete tree-level S-matrix of maximally supersymmetric Yang-Mills theory in four dimensions. Beyond the early calculations of~\cite{Witten:2003nn,Roiban:2004yf,Witten:2004cp}, considerable supporting evidence for this view has emerged in a series of recent papers~\cite{Dolan:2009wf,Bullimore:2009cb,ArkaniHamed:2009dg,Nandan:2009cc,Dolan:2010xv,Bourjaily:2010kw} that relate the twistor-string to the Grassmannian contour integral of~\cite{ArkaniHamed:2009dn}. In~\cite{Dolan:2009wf,Bullimore:2009cb,ArkaniHamed:2009dg} the twistor-string was shown to define a natural cycle in the Grassmannian by the Veronese embedding. Using contour deformation arguments, this relationship was exploited to prove that the twistor-string correctly computes all NMHV tree amplitudes~\cite{ArkaniHamed:2009dg,Nandan:2009cc} and all gluonic split helicity\footnote{A \emph{split helicity} partial amplitude has external helicity configuration $\la++\cdots+--\cdots-\ra$.} partial amplitudes~\cite{Dolan:2010xv}. Most recently, Bourjaily {\it et al.}~\cite{Bourjaily:2010kw} have expressed the twistor-string cycle for arbitrary numbers of particles and arbitrary MHV degree in terms intrinsic to the Grassmannian. The resulting integral is then conjectured to be independent of certain parameters $t^j_\ell$ that interpolate between the twistor-string integrand at $t^j_\ell=1$ and the BCFW expansion at $t^j_\ell=0$.

The relation to the Grassmannian is certainly very elegant and doubtless has much more to teach us. However, precisely because they take place in an auxiliary space -- G$(k,n)$ -- these calculations do not illuminate what is happening to the \emph{string} itself as it undergoes BCFW recursion. For example, in~\cite{ArkaniHamed:2009dg,Bourjaily:2010kw} the interpretation of the integrand for generic $t^j_\ell\in(0,1)$ is still somewhat mysterious. In consequence, the relation between the string worldsheet and the BCFW expansion is not made clear and the extent to which the twistor-string, as opposed to the Grassmannian, really knows about BCFW remains obscure.

This is not merely a matter of technical complexity. For example, twistor-string theory teaches us to expect that NMHV tree amplitudes are supported on degree 2 rational curves in twistor space. However, upon solving the BCFW recursion in momentum space~\cite{Drummond:2008cr}, one finds that the NMHV amplitudes may be written as a product of the MHV tree with a sum of certain dual superconformal invariants $R_{n;ab}$. In twistor space, each such term is supported when the external twistors lie on \emph{three} intersecting lines in a configuration of genus \emph{one}, as shown in figure~\ref{fig:3mb}. More generally\cite{Korchemsky:2009jv,Bullimore:2009cb}, each term in the BCFW decomposition of an $n$-particle N$^p$MHV tree amplitude has support when the external twistors lie on $2p+1$ lines that form a highly degenerate curve of genus $p$, and not the degree $p+1$ rational curve predicted by twistor-strings.  I emphasise that the same, high genus configurations are found whether one transforms the momentum space answer or localises on residues in the Grassmannian formula of~\cite{ArkaniHamed:2009dn}. 

\FIGURE[t]{
	\includegraphics[width=86mm]{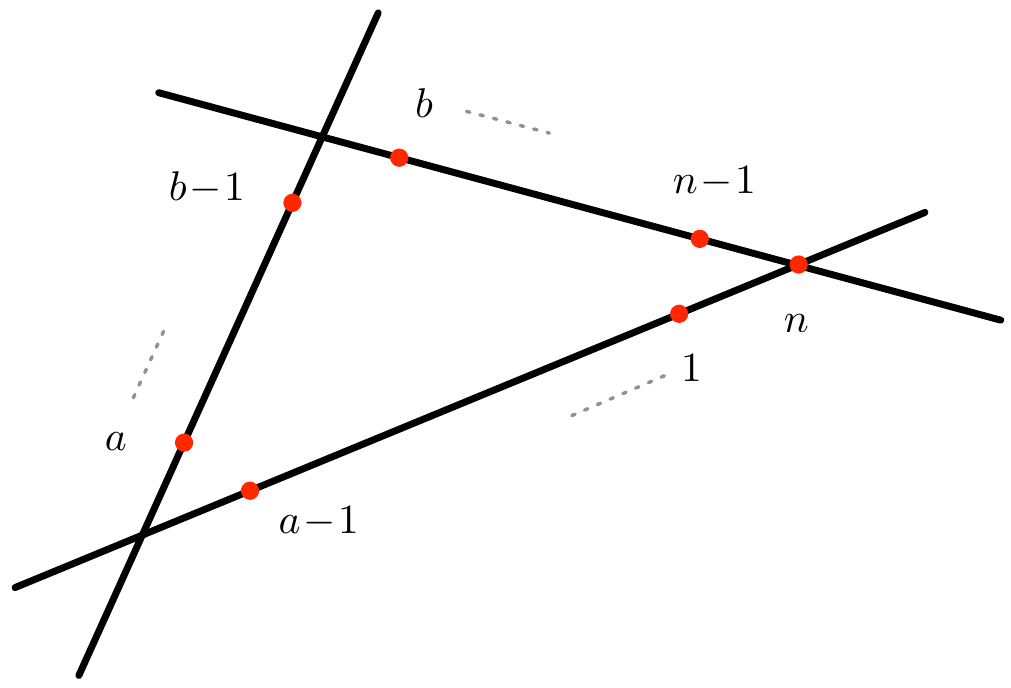}
	\caption{{\it Each term in the BCFW decomposition of an $n$-particle NMHV tree amplitude has support on three 		intersecting lines in twistor space. This is a genus one curve of degree three, and not a degree two rational 			curve as predicted by twistor-string theory.}}
	\label{fig:3mb}
}

In~\cite{Bullimore:2009cb}, these higher genus configurations were interpreted as treating each term in the BCFW expansion as a leading singularity of a multi-loop amplitude in a momentum space channel that can readily be deduced from the twistor support. This interpretation is an natural extension of the arguments of~\cite{Britto:2004ap}, where the recursion relation was originally obtained by considering a particular class of leading singularities at one loop. But, if we take the legitimate view that BCFW recursion is a relation purely among tree amplitudes, how can the $g=0$ twistor-string know about such apparently bizarre configurations as in figure~\ref{fig:N2MHVchannel} at N$^2$MHV? Certainly, there are no maps from a genus zero worldsheet to a higher genus image.

\FIGURE[t]{
	\includegraphics[width=86mm]{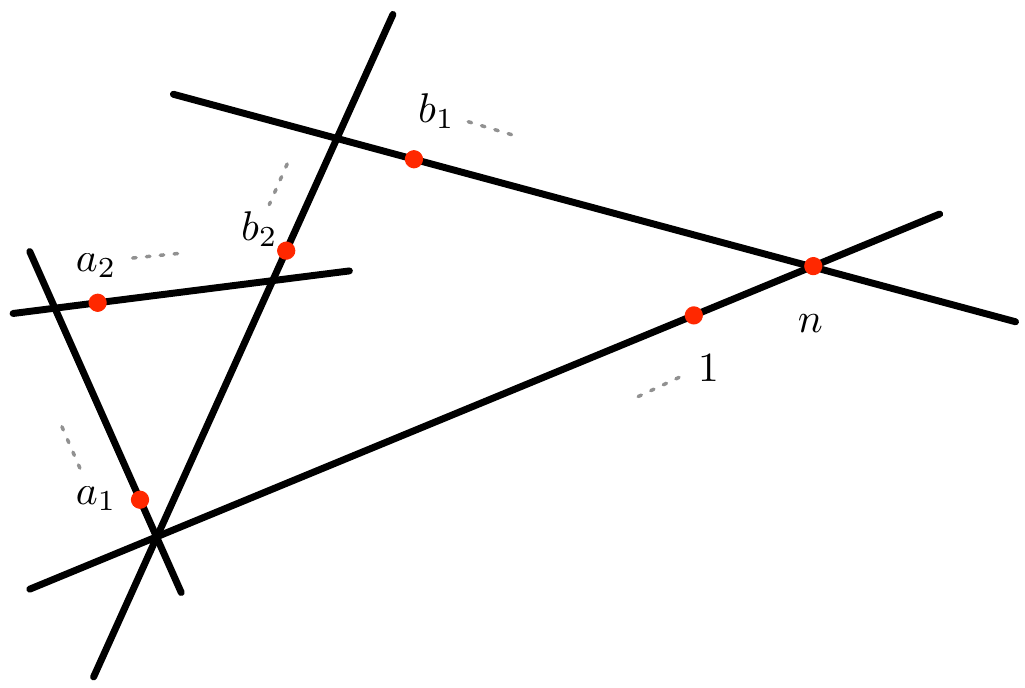}
	\caption{{\it An example of the twistor support of a summand in the solution of the BCFW relation for N$^2$MHV 		amplitudes ($A_{\rm MHV}R_{n;a_1b_1}R_{n;b_1a_1;a_2b_2}$ in the notation of~\cite{Drummond:2008cr}). The 		two triangles lie in different planes.}}
	\label{fig:N2MHVchannel}
}

This paper answers such questions. After quickly reviewing the relevant aspects of twistor-string theory in section~\ref{sec:twsamps},  in sections~\ref{sec:shift} and~\ref{sec:recursion} we derive the BCFW recursion relation directly from the twistor-string path integral. The argument closely parallels the momentum space derivation of Britto, Cachazo, Feng and Witten themselves~\cite{Britto:2005fq}, and provides the first complete proof that the leading trace contribution to the $g=0$ twistor-string path integral indeed correctly computes all tree amplitudes in $\cN=4$ SYM. Aside from the basic simplicity of the derivation -- requiring little more than an integration by parts -- one of the main advantages of this direct approach is that the worldsheet and its image in twistor space remain clearly in view at every stage. This allows us to provide a beautiful geometric interpretation of BCFW recursion for twistor-strings and in section~\ref{sec:geometry} we explain how this is compatible with the dual (tree {\it vs.} leading singularity) interpretations of the recursion.

%%%%%%%%%%%%%%%%%%%%%%%%%%%%%%%%%%%%%%%%%%%%%%%%%
%%%%%%%%%%%%%%%%%%%%%%%%%%%%%%%%%%%%%%%%%%%%%%%%%

\section{Scattering Amplitudes from Twistor-Strings}
\label{sec:twsamps}

To set the stage, let us briefly review the salient features of the twistor-string (further details can be found in~\cite{Witten:2003nn,Berkovits:2004hg,Berkovits:2004jj,Mason:2007zv,Dolan:2007vv}). Twistor-strings may be formulated starting from the action
\be
	S=\int_\Sigma Y_I\bar{D} Z^I +\  \cdots
\label{action}
\ee
where $Z:\Sigma\to\mathbb{PT}$ is a map from a compact Riemann surface $\Sigma$ to twistor space\footnote{I will denote the neighbourhood of a line in $\CP^{3|4}$ by $\PT$.}, whose homogeneous coordinates are represented by four bosonic and four fermionic worldsheet scalar fields $Z^I(\sigma)=(Z^\alpha(\sigma),\psi^a(\sigma))$. Since homogeneous coordinates are defined only up to local rescalings, these fields are valued in a holomorphic line bundle $\cL\cong Z^*\cO(1)$ and we let $\bar D$ be the (0,1)-part of a connection on $\cL$. The field $Y\in \Omega^{(1,0)}(\Sigma,\C^{4|4}\times\cL^{-1})$ is a Lagrange multiplier that forces the map to be holomorphic, so that on-shell $Z\in \mathbb{C}^{4|4}\times H^0(\Sigma,\cL)$.

There are thus $4h^0(\Sigma,\cL)$ fermionic zero-modes, which must be absorbed by vertex operator insertions if the path integral is not to vanish. For external states in $\cN=4$ SYM, the vertex operators are constructed by coupling elements
\be
	a\in H^1(\mathbb{PT},{\rm End} E)
\ee
to a worldsheet current algebra that is left implicit in~\eqref{action}. Such cohomology classes represent linearised $\cN=4$ YM supermultiplets via the Penrose transform. Expanding in the fermions,
\be
	a(Z,\psi) = g^-(Z) + \psi^a\Gamma_a(Z) + \cdots + (\psi)^4g^+(Z)
\ee
we see that $n$-particle N$^{k-2}$MHV amplitudes\footnote{Twistor and spinor conventions used in this paper are detailed in appendix~\ref{app:conventions}.} receive contributions only from the path integral sector where $h^0(\Sigma,\cL)=k$. The Riemann-Roch theorem states that for genus $g$ worldsheet
\be
	h^0(\Sigma,\cL)- h^1(\Sigma,\cL) = d +1-g
\label{RiemRoch}
\ee
where $d={\rm deg}\,\cL$ is the degree of the map, while the Kodaira vanishing theorem ensures that generically (and always at genus zero) $h^1(\Sigma,\cL)$ vanishes. Thus, at tree-level, N$^{k-2}$MHV amplitudes are supported on holomorphic curves in twistor space of degree $d=k-1$.

To compute scattering amplitudes, one first integrates out the current algebra. At leading order in the colour trace, this yields a contribution
\be
	\prod_{i=1}^n \frac{(u_i\,{\rm d}u_i)}{(u_i\,u_{i+1})}
\label{props}
\ee
summed over non-cyclic permutations of the labellings. Here, $u^a$ are homogeneous coordinates on the $\CP^1$ worldsheet, with $(u_1\,u_2)\equiv \epsilon_{ab}u_1^au_2^b$. The individual factors in~\eqref{props} may be thought of as propagators linking two points that are adjacent in some colour-ordering, and indeed they arise this way if the worldsheet current algebra comes from a holomorphic free fermion model.

Together with the cohomology classes obtained from the vertex operators, this current correlator is integrated over the moduli space $\overline{M}_{0,n}(\PT,d)$ of degree $d$ stable maps from a genus zero worldsheet with $n$ marked points to twistor space, considered up to worldsheet automorphisms. After fixing the SL$(2,\C)$ transformations, the current correlator and zero modes\footnote{The non-zero modes provide a section of a determinant line bundle that may be chosen to be trivial and have a flat (Quillen) connection, at least at genus zero. See {\it e.g.}~\cite{Mason:2007zv}.} of $Z$ (describing the holomorphic map) provide a top holomorphic form on this space and, since the integrand depends on the coordinates of $\overline{M}_{0,n}(\PT,d)$ holomorphically, we treat the path integral as a contour integral. In split signature, the appropriate choice of contour comes from demanding the map is equivariant with respect to real structures on $\Sigma$ and on $\PT$ that fix an equator and an $\RP^{3|4}$ real slice, respectively.

The twistor-string formalism makes manifest a number of important properties of the colour-ordered amplitude, such as the dihedral symmetry in the external states, and the photon decoupling identity. It is also straightforward to show that the twistor-string formula has the correct collinear and soft limits~\cite{Roiban:2004yf} as well as the correct behaviour under parity transformations~\cite{Witten:2004cp}. As it is based on a holomorphic map to a homogeneous space (whose Chern classes vanish due to $\cN=4$ supersymmetry), the twistor-string is also expected to possess a Yangian symmetry~\cite{Witten:2003nn}.

\bigskip

In the remainder of this paper, we will consider scattering amplitudes for external states that are `localised at definite twistors'. In a Dolbeault framework, these are obtained by choosing external wavefunctions
\be
\begin{aligned}
	a_i(Z)=\bar\delta^{3|4}(Z_i,Z)
	&\equiv\int_{\C^*}\frac{{\rm d}s}{s}\,\bar\delta^{4|4}(Z_i+sZ)\\
	&\equiv\int_{\C^*}\frac{{\rm d}s}{s}\,
	\bigwedge_{\alpha=1}^4\delbar\left(\frac{1}{Z_i^\alpha+sZ^\alpha}\right)\prod_{a=1}^4(\psi^a_i+s\psi^a)\,.
\end{aligned}
\label{deltabardef}
\ee
Except for the three-particle $\MHVbar$ amplitude, we will also make the genericity assumption that no two external twistors coincide in $\PT$, {\it i.e.} that $Z^I_i\neq r Z^I_j$ for any $r\in\C^*$. Note that ${\rm D}^{3|4}X\wedge\bar\delta^{3|4}(X,Y)$ may be thought of as a representative of the Poincar{\'e} dual of the diagonal $\Delta\subset\PT_X\times\PT_Y$, and that $\bar\delta^{3|4}(X,Y)$ is a distribution-valued (0,3)-form on $\PT_X\times\PT_Y$. The amplitude $A(Z_1,\ldots,Z_n)$ constructed from these wavefunctions is interpreted as a $(0,2n)$-form with compact support on $\otimes_{i=1}^n\, \PT_i$, and may be paired with classes in $H^{0,1}(\PT_i,\ \cdot\ )$ to obtain the scattering amplitude
of finite-norm wavefunctions.

%%%%%%%%%%%%%%%%%%%%%%%%%%%%%%%%%%%%%%%%%%%%%%%%%
%%%%%%%%%%%%%%%%%%%%%%%%%%%%%%%%%%%%%%%%%%%%%%%%%

\section{The BCFW Shift on Twistor Space}
\label{sec:shift}

The BCFW procedure starts by deforming the on-shell supermomenta $(\tilde\pi_A,\pi_{A'},\psi^a)$ of two states, say 1 and $n$, as~\cite{Britto:2005fq,Brandhuber:2008pf,ArkaniHamed:2008gz}
\be
	\pi_1\to\pi_1+t\pi_n\,\qquad
	\tilde\pi_n\to \tilde\pi_n-t\tilde\pi_1\, ,\qquad
	\eta_n\to\eta_n-t\eta_1\,,
\label{BCFWshiftmom}
\ee
where $t$  parametrises the deformation. (I suppress spinor and R-symmetry indices.) These shifts are generated by the vector
\be
	\pi_n\frac{\del}{\del\pi_1}-\tilde\pi_1\frac{\del}{\del\tilde\pi_n}
	-\eta_1\frac{\del}{\del\eta_n}\,,
\label{BCFWgenerator}
\ee
that acts on the $n$-particle on-shell momentum superspace. 

The correspondence between twistor space $\PT$ and complexified space-time is encoded in the incidence relations
\be
	\omega^A=\im x^{AA'}\pi_{A'}\qquad\qquad \psi^a=\theta^{A'a}\pi_{A'}\,.
\label{incidence}
\ee
These show that space-time supertranslations $(x,\theta)\to(x+a,\theta+\xi)$ are generated on $\PT$ by the vector fields $\im\pi_{A'}\del/\del\omega^A$ and $\pi_{A'}\del/\del\psi^a$. Comparing these to the momentum operator $P_{AA'} = \tilde\pi_A\pi_{A'}\leftrightarrow -\im{\del}/{\del x^{AA'}}$ and the (chiral) supersymmetry generator\footnote{See appendix~\ref{app:conventions} for details of our conventions.} $Q_{aA'}=\eta_a\pi_{A'}\leftrightarrow \del/\del\theta^{aA'}$ shows that we have a correspondence
\be
 	\tilde\pi_A\leftrightarrow \frac{\del}{\del\omega^A}
	\qquad\qquad
	\omega^A\leftrightarrow-\frac{\del}{\del\tilde\pi_A}
\label{pi-tilde-omega}
\ee
and
\be
	\eta_a\leftrightarrow\frac{\del}{\del\psi^a}
	\qquad\qquad
	\psi^a\leftrightarrow-\frac{\del}{\del\eta_a}\,.
\label{eta-psi}
\ee
In split signature space-time this obtained from treating $(\tilde\pi_A,\eta_a)$ and $(\omega^A,\psi^a)$ as Fourier conjugate variables\footnote{The absence of factors of $\im$ is accounted for by noting that if the split signature incidence relation is taken to be~\eqref{incidence} then the components of $x^{AA'}$ must be purely imaginary.}  as in~\cite{Witten:2003nn}, but note that~\eqref{pi-tilde-omega}-\eqref{eta-psi} hold even in the complex. Under this replacement, the BCFW shift is generated on twistor space by 
\be
	\omega_n\frac{\del}{\del\omega_1}+
	\pi_n\frac{\del}{\del\pi_1}+
	\psi_n\frac{\del}{\del\psi_1} = Z_n\frac{\del}{\del Z_1}\,,
\ee
so that the entire supertwistor $Z_1$ is translated towards $Z_n$, while all other external twistors are unaltered. Indeed, recalling that the shift parameter $t\in\C\cup\{\infty\}$, this Riemann sphere finds a geometric realisation as the holomorphic line $[Z_1\wedge Z_n]$ in twistor space, with $t=0$ the point $Z_1$ and $t=\infty$ the point $Z_n$.

Correspondingly, if in split signature the twistor amplitude is written as the Fourier transform of the momentum space amplitude\footnote{Here and below, the Fourier transforms are taken supersymmetrically, so that in particular $\exp(\im\omega\cdot\tilde\pi)\equiv \exp(\im\omega^A\tilde\pi_A+\im\psi^a\eta_a)$.} 
\be
	A(Z_1,\ldots,Z_n) = \int \prod_{i=1}^n \rd^{2|4}\tilde\pi_i\ \e^{\im \omega_i\cdot\tilde\pi_i}\ 
	\cA(\{\tilde\pi_1,\pi_1,\eta_1\};\ldots;\{\tilde\pi_n,\pi_n,\eta_n\})\,,
\ee
then by absorbing the translation of $\tilde\pi_n$ into the integration measure, the shifted momentum space amplitude may be transformed as
\be
\begin{aligned}
	&\int\prod_{i=1}^n \rd^{2|4}\tilde\pi_i\  \e^{\im\omega_i\cdot\tilde\pi_i}\ 
	\cA(\{\tilde\pi_1,\pi_1+t\pi_n,\eta_1\};\ldots;\{\tilde\pi_n-t\tilde\pi_1,\pi_n,\eta_n-t\eta_1\})\\
	=\ & \int\prod_{i=1}^n \rd^{2|4}\tilde\pi_i\  \e^{\im\omega_i\cdot\tilde\pi_i}\ \e^{\im t\omega_n\cdot\tilde\pi_1}\ 
	\cA(\{\tilde\pi_1,\pi_1+t\pi_n,\eta_1\};\ldots;\{\tilde\pi_n,\pi_n,\psi_n\})\\
	=\ & A(Z_1+tZ_n,Z_2,\ldots,Z_n)\,.
\end{aligned}
\ee
As $t$ varies, we obtain a one parameter family of twistor amplitudes depending on the $n-1$ twistors $Z_2,\ldots,Z_n$ and also on the line through $Z_1$ and $Z_n$. Note that the shift is meaningless for the 3-particle $\MHVbar$ amplitude, since all three external twistors coinicde. For MHV amplitudes, the line $[Z_1\wedge Z_n]$ coincides with the image of the worldsheet, but for all higher MHV levels the worldsheet does not wrap this line.

\bigskip

This argument was noted by Mason and the author in~\cite{Mason:2009sa}, but was not exploited there. Instead, that paper simply translated the right hand side of the momentum space BCFW expansion into twistorial language. Here, we will make use of the above twistor space BCFW shift to derive a recursion relation for the twistor-string directly.

%%%%%%%%%%%%%%%%%%%%%%%%%%%%%%%%%%%%%%%%%%%%%%%%%
%%%%%%%%%%%%%%%%%%%%%%%%%%%%%%%%%%%%%%%%%%%%%%%%%

\section{The Recursion Relation}
\label{sec:recursion}

We will model our proof of the twistor-string recursion relation on the proof of BCFW themselves~\cite{Britto:2005fq}. So we write the desired amplitude as
\be
	A(Z_1,Z_2,\ldots,Z_n) = \frac{1}{2\pi\im}\oint \frac{{\rm d}t}{t}\,A(Z_1+tZ_n,Z_2,\ldots,Z_n)\,,
\label{BCFWcontour}
\ee
where the contour is a small circle around $t=0$, and seek to evaluate the same integral by instead summing over contributions from the deformed amplitude, as computed from twistor-string theory.

We immediately notice that there is no contribution at $t=\infty$, for using projective invariance to rescale $Z_1+tZ_n\simeq t^{-1}Z_1+Z_n$, any such term would have to map both the first and $n^{\rm th}$ worldsheet marked points to the same point $Z_n\in\PT$. Beyond $d=0$, there are no such rational maps. (In making this argument, it is important that we used the supersymmetric BCFW shift so that the entire supertwistor $Z_1(t)$ coincides with $Z_n$ at $t=\infty$.) 

%%%%%%%%%%%%%%%%%%%%%%%%%%%%%%%%%%%%%%%%%%%%%%%%%
%%%%%%%%%%%%%%%%%%%%%%%%%%%%%%%%%%%%%%%%%%%%%%%%%

\subsection{A Contour Deformation in the Path Integral}
\label{sec:contour}

We must now understand the $t$-dependence of the amplitude itself.  From the point of view of twistor-strings, the shift affects neither the worldsheet fields $Z(u)$, nor the location of the vertex operator insertions on $\Sigma$ -- indeed it cannot, since these are integrated out in computing the amplitude. The only place the shift enters is in the vertex operator associated to the first marked point:
\be
	\bar\delta^{3|4}(Z_1,Z(u_1))\to\bar\delta^{3|4}(Z_1+tZ_n,Z(u_1))\,.
\label{shiftvertexop}
\ee
Our strategy is thus to evaluate the right hand side of~\eqref{BCFWcontour} by exchanging the order of the path integral and the $t$ contour integral.

To do this, first recall that $\bar\delta^{3|4}(X,Y)$ is defined explicitly by
\be
	\bar\delta^{3|4}(X,Y) \equiv \int_{\C^*} \frac{{\rm d}s}{s}\, \bar\delta^{4|4}(X+sY) \equiv
	\int_{\C^*}\frac{{\rm d}s}{s}\,\bigwedge_{\alpha=1}^4
	\delbar\left(\frac{1}{X^\alpha+sY^\alpha}\right)\prod_{a=1}^4(\psi^a_X+s\psi^a_Y)\,.
\label{d34defined}
\ee
In treating $\bar\delta^{3|4}(Z_i,Z(u_i))$ as a vertex operator, one of the $\delbar$ operators in~\eqref{d34defined} should be taken to act on the worldsheet coordinates. Indeed, this is explicitly found to be the case when computing MHV amplitudes -- one component of this $\bar\delta^{3|4}$-function is used to fix the integration over the worldsheet coordinate of the $i^{\rm th}$ marked point in terms of the external spinor $\pi_i$, transforming the current correlator denominator into the standard Parke-Taylor denominator.

More generally, if the external wavefunctions are represented by Dolbeault (0,1) cohomology classes on twistor space, then vertex operators are constructed by using the $n$ evaluation maps
\be
\begin{aligned}
	{\rm ev}_i: &\ \overline{M}_{0,n}(\PT,d) &\to&\quad \PT\\
	&\ (\Sigma;p_1,\ldots,p_n; Z) &\mapsto&\quad  Z(p_i)
\end{aligned}
\ee 
to pull these cohomology classes on $\PT$ back to cohomology classes on $\overline{M}_{0,n}(\PT,d)$, over which the path integral is taken. To spell this out, in the context of~\eqref{d34defined} one has
\be
\begin{aligned}
	\bar\delta^{3|4}(Z_i,Z(u_i)) &\equiv{\rm ev}_i^*\left[\bar\delta^{3|4}(Z_i,Z)\right]\\
	&=\int\frac{{\rm d}s}{s}\ {\rm ev}^*_i\!\left[\bigwedge_{\alpha=1}^4\delbar\left(\frac{1}{Z_i^\alpha+sZ^\alpha}\right)
	\prod_{a=1}^4(\psi_i^a + s\psi^a)\right]\\
	&=\int\frac{{\rm d}s}{s}\,\bigwedge_{\alpha=1}^4\delbar\left(\frac{1}{Z_i^\alpha+sZ^\alpha(u_i)}\right)	\prod_{a=1}^4(\psi_i^a + s\psi^a(u_i))\,,
\end{aligned}
\label{vertexpullback}
\ee
where in going to the last line, we use the fact that the $\delbar$-operator commutes with pullbacks (the evaluation map is holomorphic) and replaced the homogeneous coordinates $Z^I$ by their pullback $Z^I(u_i)={\rm ev}_i^*(Z^I)$. The important point is that because we commuted it through this pullback, the $\delbar$-operator that appears in the last line of~\eqref{vertexpullback} is really\footnote{More accurately, $\bar\delta^{3|4}(Z_i,Z(u_i))$ is a (0,3) form on $\PT_i\times \overline{M}_{0,n}(\PT,d)$ and so contains a sum of terms that are $(0,p)$-forms on $\PT_i$ and $(0,3-p)$-forms on the moduli space. The vertex operator is formed from the summand that is a (0,1)-form on the moduli space.}  the $\delbar$-operator on $\overline{M}_{0,n}(\PT,d)$.

Now, integrating by parts, we can remove one of these $\delbar$-operators from the $t$-dependent vertex operator and have it act on the remaining terms in the path integral\footnote{Equivalently, in a {\v C}ech framework this may be thought of as a global residue theorem on the moduli space.}. This leaves us with a `bare' factor of $1/(Z_1+tZ_n+sZ(u_1))^\alpha$ for some component $\alpha$, which can be used to perform the $t$ contour integral. In fact, it will not be necessary to do this $t$ integral explicitly. We first recast the BCFW contour integral~\eqref{BCFWcontour} in Dolbeault form as
\be
\begin{aligned}
	A(Z_1,\ldots,Z_n) &= \int {\rm d}t\wedge\delbar\left(\frac{1}{t}\right)\, A(Z_1+tZ_n,Z_2,\ldots,Z_n)\\
		&= -\int \frac{{\rm d}t}{t}\wedge\delbar\left[A(Z_1+tZ_n,Z_2,\ldots,Z_n)\right]\,.
\end{aligned}
\label{intbypartst}
\ee
Having transferred a $\delbar$-operator off one of the components of the shifted vertex operator above, in the second line of~\eqref{intbypartst} we promptly put one back on again, but now the $\delbar$-operator associated with the BCFW shift parameter. Equivalently, the contour that used to wrap around $t=0$ now wraps around $(Z_1+tZ_n+sZ(u_1))^\alpha=0$ and treating this as a simple pole in $t$ has the same localising effect as the original Dolbeault $\bar\delta$-function.

At this stage, the BCFW shift parameter is really no different from any of the other scaling parameters in the vertex operators. Indeed, we may define~\cite{Mason:2009sa}
\be
	\bar\delta^{2|4}(Z_1,Z_2,Z_3)\equiv\int_{(\C^*)^2}\frac{{\rm d}s}{s}\frac{{\rm d}t}{t}\,\bar\delta^{4|4}(Z_1+sZ_2+tZ_3)
\ee
that forces its three arguments to be collinear in projective twistor space. Thus, after our integrations by parts, while  the marked points $\{2,3,\ldots,n\}$ must still be mapped to some fixed points $\{Z_2,\ldots,Z_n\}\in\PT$,  the first marked point is only required to map to the line $[Z_1\wedge Z_n]$.

%%%%%%%%%%%%%%%%%%%%%%%%%%%%%%%%%%%%%%%%%%%%%%%%%
%%%%%%%%%%%%%%%%%%%%%%%%%%%%%%%%%%%%%%%%%%%%%%%%%

\subsection{Localising on Boundary Divisors}
\label{sec:localisation}

Having peeled it off from the vertex operator, we must now evaluate the action of the moduli space $\delbar$-operator on the remaining terms in the path integrand.

This operator does not see the remaining vertex operators, since these represent $\delbar$ cohomology classes. It also does not see the measure on the space of zero modes of the worldsheet map $Z(u)$, since this measure is holomorphic (even Calabi-Yau in our $\cN=4$ context). Thus, the only potential contribution is
\be
	\delbar\left[\frac{1}{{\rm vol(SL}(2,\C))}\prod_{i=1}^n\frac{(u_i\,{\rm d}u_i)}{(u_i\,u_{i+1})}\right]
\label{dbarcurrcorr}
\ee
coming from the current correlator, considered up to worldsheet automorphisms. After quotienting by SL$(2,\C)$ transformations, this current correlator is an $(n-3,0)$-form on the moduli space $\overline{M}_{0,n}$ of genus zero $n$-pointed stable curves, parametrising the locations of the vertex operators up to SL$(2,\C)$ equivalence\footnote{For $n\geq3$ there is a map from $\overline{M}_{0,n}(\PT,d)\to\overline{M}_{0,n}$ that forgets about the map to the target space ({\it i.e.}, ignores the behaviour of the worldsheet field $Z(u)$). The current correlator is really pulled back to $\overline{M}_{0,n}(\PT,d)$ by this map, and we again commute the $\delbar$-operator on $\overline{M}_{0,n}(\PT,d)$ through this pullback to act on the current correlator as the $\delbar$-operator on $\overline{M}_{0,n}$.}. Clearly, the correlator has no poles on the dense open set $M_{0,n}\subset\overline{M}_{0,n}$ describing $n$ \emph{distinct} points on $\CP^1$, so the only possible contributions to~\eqref{dbarcurrcorr} come from the boundary $\overline{M}_{0,n}\backslash M_{0,n}$. In the Deligne-Mumford compactification (standard in string theory), the limiting configurations where marked points collide are described in terms of nodal worldsheets with the marked points distributed among the different  irreducible components in a way that reflects the way the collision is approached.

In fact, this story is very familiar in string theory. From the worldsheet point of view, the $\delbar$-operator on $\overline{M}_{0,n}(\PT,d)$ started out life as the antiholomorphic BRST operator corresponding to the scalar supercharge of a twisted $(0,2)$-model~\cite{Mason:2007zv}. The path integrand is BRST closed, except for possible contributions from the boundary of the moduli space where the worldsheet degenerates (see {\it e.g.}~\cite{Bershadsky:1993cx}).

\FIGURE[t]{
	\includegraphics[width=150mm]{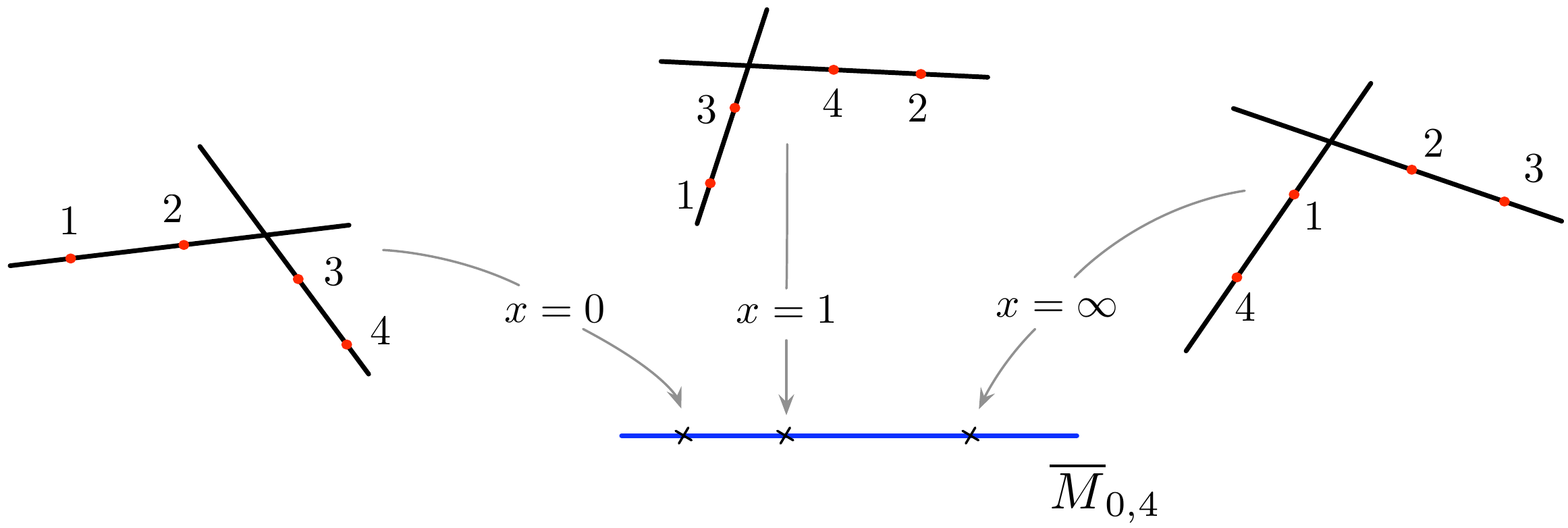}
	\caption{{\it The compactified moduli space of a four-pointed rational curve is a copy of $\CP^1$ shown here as a 				(blue) complex line. The three points $x=0$, 1 and $\infty\in\overline{M}_{0,4}$ correspond to the three 				possible stable worldsheet degenerations.}}
	\label{fig:M04}
}

\bigskip

To examine the behaviour of the correlator near $\overline{M}_{0,n}\backslash M_{0,n}$, let us first consider the simplest case of $n=4$. As is well-known, $M_{0,4}\cong\CP^1-\{0,1,\infty\}$, parametrising the possible location of the fourth marked point once three are fixed, while the compactification $\overline{M}_{0,4}\cong\CP^1$ with three special points $\{0,1,\infty\}$ corresponding to the three possible stable\footnote{A curve is \emph{stable} if it has at most a finite automorphism group. At genus zero this implies that each curve component contains at least three special points (either  marked points or nodes).} worldsheet degenerations. The unique independent cross-ratio
\be
	x\equiv\frac{(u_1\,u_2)(u_3\,u_4)}{(u_4\,u_1)(u_2\,u_3)}
\ee
provides a local coordinate on $\overline{M}_{0,4}$ with the points $x=\{0,1,\infty\}$ corresponding to the nodal curves  shown in figure~\ref{fig:M04}. Note that two of these boundary divisors --  $x=0$ and $x=\infty$ -- are compatible with the cyclic ordering of the marked points, while the remaining divisor $x=1$ is not.

The $n=4$ current correlator is naturally written in terms of this cross-ratio. For example, using the SL(2,$\C$) invariance to fix $u_1$, $u_3$ and $u_4$ to some arbitrary values, we have
\be
	\frac{1}{{\rm vol (SL}(2,\C))}\prod_{i=1}^4 \frac{(u_i\,{\rm d}u_i)}{(u_i\,u_{i+1})} 
	= \frac{(u_1\,u_3)(u_2\,{\rm d}u_2)}{(u_1\,u_2)(u_2\,u_3)}
	 = \frac{(u_1\,{\rm d}u_2)}{(u_1\,u_2)} - \frac{({\rm d}u_2\,u_3)}{(u_2\,u_3)}
	 = \frac{{\rm d}x}{x}\,.
\label{currcorr4}
\ee
(Any other choice of gauge fixing leads to the same result.) This makes it clear that the current correlator has a simple pole at $x=\{0,\infty\}\in\overline{M}_{0,4}$ -- {\it i.e.} on each boundary divisor that is compatible with the cyclic ordering -- while it remains regular at $x=1$. Thus, when the $\delbar$-operator acts on the four particle current correlator, we obtain a sum of two terms, each one localising on the space of holomorphic maps from worldsheets that have degenerated in a way compatible with the cyclic ordering.

Focussing on the $x=0$ term, the residue of~\eqref{currcorr4} is of course $2\pi\im$, but it is instructive to express this as
\begin{multline}
	 \frac{1}{2\pi\im}\oint\frac{{\rm d}x}{x}=1\\
	 = \frac{1}{{\rm vol(SL}(2,\C))} \frac{(u_1\,{\rm d}u_1)(u_2\,{\rm d}u_2)(u_*{\rm d}u_*)}{(u_*\,u_1)(u_1\,u_2)(u_2\,u_*)}
	\times
	\frac{1}{{\rm vol(SL}(2,\C))} \frac{(v_*\,{\rm d}v_*)(v_3\,{\rm d}v_3)(v_4\,{\rm d}v_4)}{(v_*\,v_3)(v_3\,v_4)(v_4\,v_*)}\,,
\label{residue4}
\end{multline}
where $u$ and $v$ are homogeneous coordinates on the two irreducible components of the wordsheet (each of which is a $\CP^1$), and $u_*$, $v_*$ are the coordinates of the node on each component. Using the freedom to make independent SL$(2,\C)$ transformations on each component, we can fix the three special points on each component to some arbitrary values and, accounting for the resulting Jacobians, the right hand side of~\eqref{residue4} is indeed unity. The reason for writing~\eqref{residue4} is to show we can interpret the residue of the 4-point current correlator as a product of lower-point current correlators on each component of the pinched worldsheet, with the nodes themselves participating in the correlator. As illustrated in figure~\ref{fig:4ptresidue}, this is exactly what we would expect from considering the original correlator as the worldsheet degenerates.

\FIGURE[t]{
	\includegraphics[width=130mm]{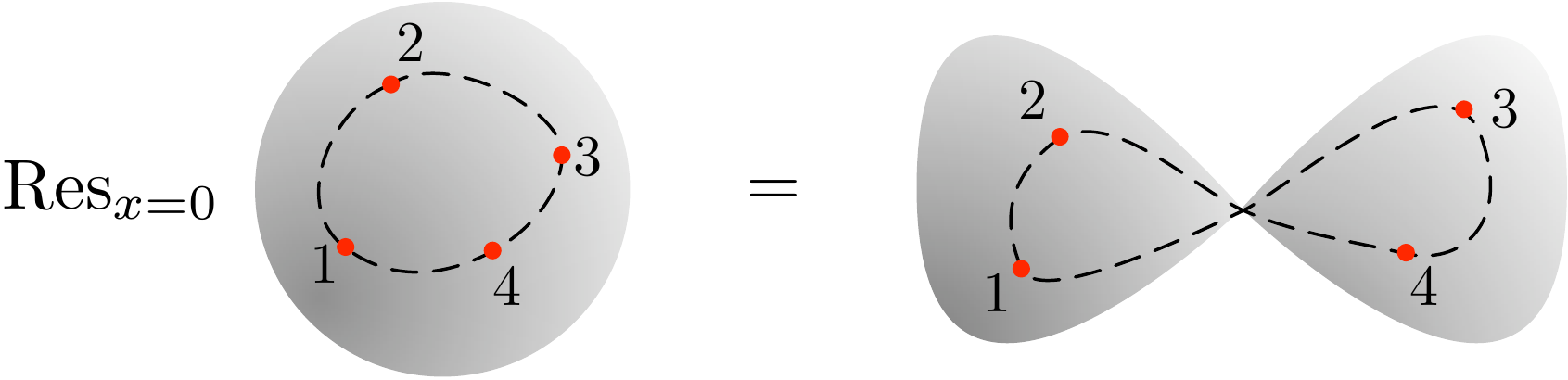}
	\caption{{\it The residue of the current correlator at the boundary of the moduli space is a product of current 		correlators on the two worldsheet components, with the node included as a new 'source'. This source arises 		because, from the perspective of either component, the two marked points on the other component have 		collided at the location of the node.}}
	\label{fig:4ptresidue}
}

Note that while~\eqref{residue4} gives the residue of the current correlator, to obtain the residue of the full path integrand we must also consider the behaviour of the map itself on these boundary divisors, and how this is affects the vertex operators. This is carried out in section~\ref{sec:BCFWnodal}. There we will see that the remaining term at $x=\infty$ actually gives no contribution as a consequence of our genericity assumption on the external twistors. Indeed, since we are considering the $Z_1\to Z_1+tZ_n$ shift, we would not expect to find a contribution from $x=\infty$, where the first and fourth $(=n^{\rm th})$ vertex operators are on the \emph{same} worldsheet component.

\bigskip

Returning to the $n$ particle case, we can describe a patch of $\overline{M}_{0,n}$ by making a choice of $n-3$ independent cross-ratios. For general $n$, there are many possible ways to choose these cross-ratios. A particularly useful choice for our purposes is 
\be	
	x_j \equiv \frac{(u_1\,u_j)(u_{j+1}\,u_n)}{(u_n\,u_1) (u_j\,u_{j+1})}\, \qquad \hbox{for}\qquad j\in\{2,3,\ldots,n-2\}\,,
\label{xratios}
\ee
the point being that, as in~\eqref{currcorr4}, the $n$-point current correlator becomes simply
\be
	\frac{1}{{\rm vol (SL}(2,\C))}\prod_{i=1}^n \frac{(u_i\,{\rm d}u_i)}{(u_i\,u_{i+1})}
	= \bigwedge_{j=2}^{n-2}\, \frac{{\rm d}x_j}{x_j}\,,
\label{xmeasure}
\ee
where we have used the SL(2,$\C$) freedom to fix the 1$^{\rm st}$, $(n-1)^{\rm st}$ and $n^{\rm th}$ marked points.
Notice that the equality of these two expressions proves that our cross-ratios are indeed independent. Thus, when the $\delbar$-operator acts on the current correlator, in the coordinate patch covered by our choice of cross-ratios we obtain a sum of terms each localising on a boundary divisor where an individual $x_i$ vanishes.

Which divisor is this? The cross-ratio $x_i\to0$ when either the $1^{\rm st}$ and $i^{\rm th}$, or the $(i\!+\!1)^{\rm st}$ and $n^{\rm th}$ marked points approach each other. Supposing that $u_i\to u_1$, then as well as $x_i\to0$, na{\" i}vely
\be
	x_{i-1} \equiv \frac{(u_1\,u_{i-1})(u_i\,u_n)}{(u_n\,u_1)(u_{i-1}\,u_i)}
	\to \frac{(u_1\,u_{i-1})(u_1\,u_n)}{(u_n\,u_1)(u_{i-1}\,u_1)} = 1\,,
\ee
contradicting the fact that our cross-ratios are to be treated as independent in order for them to provide coordinates on the moduli space. In order for $x_{i-1}$ to remain generic as $x_i\to0$, it must be that the $(i-1)^{\rm st}$ marked point also approaches marked points $1$ and $i$ at the same rate.  Iterating the argument, we see that the boundary divisor where $x_i\to0$ with all other $x_j$ remaining unaffected corresponds to a nodal worldsheet in which all of the marked points $\{1,2,\ldots,i\}$ lie on one component and the remaining marked points $\{i+1,\ldots,n-1,n\}$ lie on the other. Note also that~\eqref{xmeasure} remains regular when any $x_i\to1$, corresponding to the $(j+1)^{\rm st}$ marked point colliding with the first, while the $j^{\rm th}$ remains generic and thus describing a boundary divisor that is not compatible with the cyclic ordering.

When $n>4$ it is less easy\footnote{One possible construction of $\overline{M}_{0,n}$ is to blow up $\CP^{n-3}$ at $n-1$ points in general position, then again along the lines joining pairs of these points, then again along the planes joining triples of points and so on, a total of $n-4$ times~\cite{Kapranov:modulicurves}. In particular, $\overline{M}_{0,4}\cong\CP^1$ and $\overline{M}_{0,5}\cong {\rm dP}(5)$.} to find global coordinates on $\overline{M}_{0,n}$ and there are other boundary divisors that are not simply at $x_j=\infty$. In fact, on account of the cyclic symmetry of the original expression, it is clear that the $n$-particle current correlator has a simple poles on every boundary divisor that is compatible with the cyclic ordering, generalising the $n=4$ case. We can exhibit these poles explicitly by choosing a different SL$(2,\C)$ gauge fixing and introducing a new set of cross-ratios $y_j$ that are related to the $x_j$ by a cyclic permutation of all $n$ marked points. These $y_j$ are then good coordinates in the neighbourhood of a boundary divisor where one of the $y_i\to0$, and it is clear that the current correlator has a simple pole here too. Once again we will find that, under the $[Z_1\wedge Z_n]$ shift, boundary divisors in which the first and $n^{\rm th}$ marked points are on the same worldsheet component do not contribute for generic external momenta.

There is another way to see why boundary divisors in which the distribution of marked points among the irreducible components does not respect the cyclic ordering do not arise as residues of~\eqref{xmeasure}. The dependence of the current correlator on the location of the $i^{\rm th}$ marked point is given by 
$$
	\frac{(u_{i-1}\,u_{i+1})(u_i\,{\rm d}u_i)}{(u_{i-1}\,u_i)(u_i\,u_{i+1})}\,,
$$
where we have included a factor of $(u_{i-1}\,u_{i+1})$ so as to obtain a homogeneous expression. This may be thought of as a meromorphic 1-form
\be
	 \omega\equiv \frac{(u_{i-1}\,u_{i+1})(u\,{\rm d}u)}{(u_{i-1}\,u)(u\,u_{i+1})}
	 \label{meroform}
\ee
on $\Sigma$, evaluated at the location of the $i^{\rm th}$ marked point.  This form is characterized by having simple poles only at the locations of the two adjacent marked points, with residues $\pm1$ (see also~\cite{Witten:1990hr}), and so $\omega$ has neither poles nor zeros on $M_{0,n}\subset\overline{M}_{0,n}$. To extend the current correlator over the boundary of $\overline{M}_{0,n}$, we must extend~\eqref{meroform} to a meromorphic form on a nodal worldsheet. Such a form is really a pair $(\omega_1,\omega_2)$ of meromorphic forms, one on each irreducible component. As well as poles at the marked points $i-1$ and $i+1$,  $\omega_1$ and $\omega_2$ are permitted to have simple poles at the nodes, again with equal and opposite residues on each component. These properties uniquely fix\footnote{As before, $u$ and $v$ are homogeneous coordinates on each irreducible worldsheet component, with $u_*$ and $v_*$ denoting the coordinates of the node.}
\be
	(\omega_1,\omega_2) 
	= \left(\frac{(u_{i-1}\,u_*)(u\,{\rm d}u)}{(u_{i-1}\,u)(u\,u_*)},\frac{(v_*\,v_{i+1})(v\,{\rm d}v)}{(v_*\,v)(v\,v_{i+1})}\right)
\ee
on boundary divisors where $i-1$ and $i+1$ lie on different curve components, and
\be
	(\omega_1,\omega_2)
	=\left(\frac{(u_{i-1}\,u_{i+1})(u\,{\rm d}u)}{(u_{i-1}\,u)(u\,u_{i+1})}, \ 0\ \right)
\ee
if $i-1$ and $i+1$ remain on the same boundary component. The reason $\omega_2=0$ in this case is simply that there are no meromorphic forms on a $\CP^1$ with a simple pole at just one point (the node). Thus, our form has a \emph{zero} on boundary divisors in $\overline{M}_{0,n}$ for which the $i^{\rm th}$ marked point is on a different curve component from both the $(i-1)^{\rm st}$ and $(i+1)^{\rm st}$ marked points. This zero cancels any potential pole on boundary divisors that are not compatible with the cyclic ordering.

Having identified the relevant boundary contributions, it is easy to compute the residue. For example, for those boundary divisors that are visible in the $x_j$ coordinates, from~\eqref{xmeasure} we immediately find
\be
	 \delbar\left(\bigwedge_{j=2}^{n-2}\frac{{\rm d}x_j}{x_j}\right)
	= \ \sum_{i=2}^{n-2} {\rm d}x_i\wedge\bar\delta(x_i)\bigwedge_{j\neq i}\frac{dx_j}{x_j}\,,
\ee
and after integrating out $x_i$ and localising on $x_i=0$, the remaining form in the $i^{\rm th}$ summand is
\begin{multline}
	\frac{1}{{\rm vol(SL}(2,\C))}
	\frac{(u_1{\rm d}u_1)\cdots(u_i{\rm d}u_i)(u_*{\rm d}u_*)}{(u_*\,u_1)\cdots(u_{i-1}\,u_i)(u_i\,u_*)}\\
	\times\ 
	\frac{1}{{\rm vol(SL}(2,\C))}
	\frac{(v_*\,{\rm d}v_*)(v_{i+1}\,{\rm d}v_{i+1})\cdots(v_n\,{\rm d}v_n)}{(v_*\,v_{i+1})(v_{i+1}\,v_{i+2})\cdots(v_n\,v_*)}\,,
\end{multline}
exactly as one expects from the picture of the current correlator becoming pinched as the worldsheet degenerates, and generalising figure~\ref{fig:4ptresidue} to the $n$-particle case in the obvious way. Including also the boundary divisors `at infinity' in the $x_j$ coordinates, we obtain a similar term for every boundary divisor that is compatible with the cyclic ordering.

%%%%%%%%%%%%%%%%%%%%%%%%%%%%%%%%%%%%%%%%%%%%%%%%%
%%%%%%%%%%%%%%%%%%%%%%%%%%%%%%%%%%%%%%%%%%%%%%%%%

\subsection{BCFW Recursion from Maps of Nodal Worldsheets}
\label{sec:BCFWnodal}

We have seen that the contour integral~\eqref{BCFWcontour} localises on a sum of terms each describing a nodal worldsheet. To obtain the recursion relation, all that remains is to understand how to describe a holomorphic map to twistor space from such nodal curves. 

This is straightforward: a map from a nodal curve is simply a pair of maps, one for each worldsheet component, together with the constraint that both maps send the node to the same point in the target space. If the original map has degree $d$, then the degrees $(d_1,d_2)$ of the individual maps must obey $d_1+d_2=d$ (as well as $d_i\geq0$). At genus zero, a degree $d$ map supports N$^{d-1}$MHV amplitudes, so this constraint corresponds to the condition that an N$^p$MHV amplitude decomposes under BCFW recursion into a sum of products of N$^{p_1}$MHV and N$^{p_2}$MHV amplitudes with $p_1+p_2 = p-1$.

Explicitly, if we write a degree $d$ map from an irreducible worldsheet as
\be
	Z(u) = \sum_{a=0}^d\, Y_a \prod_{b\neq a}\frac{(u\ u_b)}{(u_au_b)}
\ee
where $Y_a\in\PT$ are $d+1$ arbitrary reference points and $u_a$ any $d+1$ points on the worldsheet\footnote{The $u_a$ may be interpreted as the `centres' of the multi-instanton in the worldsheet gauge field and do not need to coincide with the marked points (though this is often convenient). All choices of these locations are gauge equivalent -- a holomorphic line bundle $\cL\to\CP^1$ is completely specified by its degree -- so they are not integrated over. This statement fails at higher genus.}, then the map from a degenerate worldsheet may be written as
\be
	Z(u) =\sum_{b=0}^{d_1}\, Y_b\prod_{e\neq b} \frac{(u\ u_e)}{(u_bu_e)}\,,\qquad
	Z(v) =\sum_{c=0}^{d_2}\, Y_c\prod_{f\neq c} \frac{(v\ v_f)}{(v_cv_f)}\,,
\label{reduciblemap}
\ee
with the requirement that $Z(u_*)=Z(v_*)$. The original map thus involves a choice of the $3+4d$ bosonic components of the $\{Y_a^\alpha\}$ (up to overall scaling). Accounting for both a separate rescaling of each map and for the constraint $Z(u_*)=Z(v_*)$,  \eqref{reduciblemap} likewise involves a choice of $(3+4d_1) + (3+4d_2) - 3 = 3+4d$ bosonic degrees of freedom. This just reflects the fact that the boundary components of $\overline{M}_{0,n}(\PT,d)$ are inherited from boundary components of $\overline{M}_{0,n}$.

Combining all the pieces, the BCFW contour integral~\eqref{BCFWcontour} for twistor-strings gives the relation
\be
\begin{aligned}
	&\hspace{-0.5cm}A(Z_1,Z_2,\ldots,Z_n) =\\
	&\ \sum_{i=2}^{n-2}\sum_{d_1+d_2=d}\int\frac{{\rm d}t}{t}\prod_{a=0}^{d_1}
	\frac{{\rm d}^{4|4}Y_a}{{\rm vol(GL}(2,\C))}
	\prod_{j=1}^i \frac{(u_j\,{\rm d}u_j)}{(u_{j-1}\,u_j)}\wedge\bar\delta^{3|4}(Z_j(t),Z(u_j))\\
	&\ \hspace{2.5cm}\times\frac{(u_*\,{\rm d}u_*)}{(u_i\,u_*)}\wedge 
	\bar\delta^{3|4}(Z(u_*),Z(v_*))\wedge\frac{(v_*\,{\rm d}v_*)}{(v_*\,v_{i+1})}\\
	&\ \hspace{2.5cm}\times\prod_{b=0}^{d_2}\frac{{\rm d}^{4|4}Y_b}{{\rm vol(GL}(2,\C))}
	\prod_{k=i+1}^n\frac{(v_k\,{\rm d}v_k)}{(v_k\,v_{k+1})}\wedge \bar\delta^{3|4}(Z_k,Z(v_k))\\
	&\ +\ \ldots\,,
\end{aligned}
\label{stringBCFW1}
\ee
where we set $u_0\equiv u_*$ and $v_{n+1}\equiv v_*$, and where $Z_1(t) = Z_1+tZ_n$ when $j=1$, and $Z_j(t)= Z_j$ otherwise. We also recall that the $t$ integral is here understood as being fixed by the $\bar\delta$-functions in the vertex operator. The ellipsis represents contributions from cyclic boundary divisors that have marked points 1 and $n$ on the same worldsheet component. These have an analogous form but are neglected since, as verified below, with the $[Z_1\wedge Z_n]$ shift they have singular support on momentum space. Similarly, although in principle both $(d_1,d_2)=(0,d)$ and $(d_1,d_2)=(d,0)$ are allowed in~\eqref{stringBCFW1}, the $(d,0)$ term has no support when the external twistors are distinct as we assumed (otherwise it corresponds to a collinear singularity of the amplitude), so that only the `left' subamplitude may be $\MHVbar$.

The factor of $\bar\delta^{3|4}(Z(u_*),Z(v_*))$ in~\eqref{stringBCFW1} enforces the constraint that our map indeed maps the node on each branch of the worldsheet to the same point in $\PT$.  If we write this as
\be
	\bar\delta^{3|4}(Z(u_*),Z(v_*)) = \int {\rm D}^{3|4}Z\wedge\bar\delta^{3|4}(Z,Z(u_*))\wedge \bar\delta^{3|4}(Z,Z(v_*))
\ee
for some auxiliary point $Z\in\PT$ (not to be confused with the map!), then the terms in~\eqref{stringBCFW1} that depend on the two worldsheet components decouple from one another. We can then recast~\eqref{stringBCFW1} as the recursion relation\footnote{It would perhaps be better to write the left subamplitude as $A(Z_1+tZ_n,Z_2,\ldots,Z^\vee)$ to emphasise that this amplitude is a (0,2)-form in each of its first $i$ slots, but only a $(0,1)$-form in the auxiliary twistor $Z$. This follows from the fact that we need to treat $\bar\delta^{3|4}(Z(u_*),Z(v_*))$ as a (0,2)-form on the `left' moduli space and a (0,1)-form on the `right' moduli space, compensating for the fact that the antiholomorphic forms in the shifted vertex operator $\bar\delta^{3|4}(Z_1+tZ_n,Z(u_1))$ are associated either with $Z_1$ or with the shift parameter. Thus the shifted subamplitude plays the role of preparing a wavefunction in $Z$ which is paired with the $Z$-dependent unshifted subamplitude.}
\be
	A(Z_1,\ldots,Z_n) 
	= \sum \int_{\C^*\times\PT}\!\frac{{\rm d}t}{t}\,{\rm D}^{3|4}Z\  A(Z_1+tZ_n,Z_2,\ldots,Z_i,Z)\ A(Z,Z_{i+1},\ldots,Z_n)\,,
\label{stringBCFW2}
\ee
where the sum is taken over $2\leq i\leq n-2$ as well as the MHV levels of the subamplitudes.

\bigskip

Readers familiar with~\cite{Mason:2009sa} will recognise~\eqref{stringBCFW2} as the twistor space form of BCFW recursion. To summarise the arguments of~\cite{Mason:2009sa}, in split signature we write
\be
\begin{aligned}
	A(Z_1+tZ_n,Z_2,\ldots,Z_i,Z) 
	&= \int {\rm d}^{2|4}\tilde\pi\ \e^{\im \omega\cdot\tilde\pi} \tilde{A}(Z_1+tZ_n,Z_2,\ldots,Z_i,\{\tilde\pi,\pi,\eta\})\\
	A(Z,Z_{i+1},\ldots,Z_n) 
	&= \int {\rm d}^{2|4} \tilde\pi'\ \e^{\im \omega\cdot\tilde\pi'} \tilde{A}(\{\tilde\pi',\pi,\eta'\},Z_{i+1},\ldots,Z_n)\,,
\end{aligned}
\ee
to replace the internal twistor by its Fourier transform. Treating ${\rm D}^{3|4}Z$ as $[\pi\,{\rm d}\pi]\, {\rm d}^{2|4}\omega$ in~\eqref{stringBCFW2}, the integral over the auxiliary $(\omega^A,\psi^a)$ gives $\delta^{2|4}(\tilde\pi+\tilde\pi')$ which can be used to perform the $\tilde\pi'$ integral. The remaining integral measure $[\pi\,{\rm d}\pi]\,{\rm d}^2\tilde\pi\,{\rm d}^4\eta$ is simply an integral over the on-shell momentum superspace of the internal particle. Taking the Fourier transform of both sides of~\eqref{stringBCFW2} with respect to all the external $(\omega_i^A,\psi^a_i)$, we find the momentum space amplitude
\be
\begin{aligned}
	&\cA(1,\ldots,n) = \\
	&\sum \int \frac{{\rm d}t}{t} [\pi\,{\rm d}\pi]\,\rd^2\tilde\pi\, \rd^4\eta\ 
	\cA(\hat 1(t),2,\ldots,i,\{\tilde\pi,\pi,\eta\})\  \cA(\{-\tilde\pi,\pi,-\eta\},i+1,\ldots,\hat n(t))\,,
\end{aligned}
\label{BCFWmom}
\ee
where $\hat{1}(t)$ and $\hat{n}(t)$ denote the usual BCFW shifted states as in~\eqref{BCFWshiftmom}, still for arbitrary values of shift parameter $t$. All the amplitudes in~\eqref{BCFWmom} involve their corresponding momentum conserving $\delta$-functions. In each summand on the right hand side, one set of these $\delta$-functions may be used to freeze the integration variables $t$ and $(\pi,\tilde\pi)$ (up to their usual scaling), ensuring
\be
	\tilde\pi\pi + \sum_{j=1}^i \tilde\pi_j\pi_j + t\tilde\pi_1\pi_n = 0\,,
\ee
which are just their standard BCFW values. The Jacobian from solving the $\delta$-functions for $(\pi,\tilde\pi,t)$ combines with the factor of $1/t$ in the integration measure to reproduce the usual $1/p_L^2$ propagator factor\footnote{The twistor space recursion relation given in~\cite{Mason:2009sa} also involved various `sign operators' or Hilbert transforms, arising from a careful treatment of certain modulus signs associated with this Jacobian. On complex twistor space, such effects are likely to be incorporated in a better cohomological understanding of the external states, and their proper relation to split signature representatives. These issues are beyond the scope of this paper, but note that it has been known since~\cite{Roiban:2004yf} that simply replacing $\bar\delta^{3|4}(Z_i,Z)$ by real $\delta$-functions in real twistors is too na{\"i}ve, for reasons associated with essentially the same modulus.} (see~\cite{Mason:2009sa} for further details).

This argument also shows that contributions from boundary divisors with 1 and $n$ on the same curve component only have support when some proper subset of the \emph{unshifted} external momenta sum to a null momentum. They correspond to factorization channels of the overall amplitude that are not shared by its BCFW expansion and are usually neglected on momentum space. Of course, these other boundary divisors are important for other choices of BCFW shift.

\bigskip

We have thus proved that the leading trace part of the genus zero twistor-string path integral obeys the BCFW recursion relation. Together with the well-known computations of the 3-point $\MHVbar$ and MHV amplitudes that seed the recursion relation, this proves that the twistor-string correctly computes all tree amplitudes in $\cN=4$ super Yang-Mills.

%%%%%%%%%%%%%%%%%%%%%%%%%%%%%%%%%%%%%%%%%%%%%%%%%
%%%%%%%%%%%%%%%%%%%%%%%%%%%%%%%%%%%%%%%%%%%%%%%%%

\section{The Geometry of BCFW Recursion}
\label{sec:geometry}

The recursion relation~\eqref{stringBCFW2} has a pleasing geometric interpretation: the subamplitude $\int {\rm d}t/t\ A(Z_1+tZ_n,Z_2,\ldots,Z_i,Z)$ is supported where the external points $Z_2,\ldots Z_i$ lie on a degree $d_1$ curve that, on account of the shifted vertex operator, must also intersect the line $[Z_1\wedge Z_n]$. The subamplitude $A(Z,Z_{i+1},\ldots,Z_n)$ is has support where  $Z_i,\ldots,Z_n$ lie on a degree $d_2$ curve. These two curves must intersect each other at some point $Z$, which may be anywhere in twistor space. As shown on the left of figure~\ref{fig:channel}, this picture reflects the momentum space sum over factorization channels of the original amplitude. The fact that a factorization channel in momentum space corresponds to an intersection in twistor space was established quite generally in~\cite{Bullimore:2009cb} (see also~\cite{Vergu:2006np}). Indeed, the channel diagram can be viewed as nothing more than the dual graph of the map, with the different subamplitudes corresponding to the different worldsheet components, and the MHV levels labelling the map degree.

\FIGURE[t]{
	\includegraphics[width=150mm]{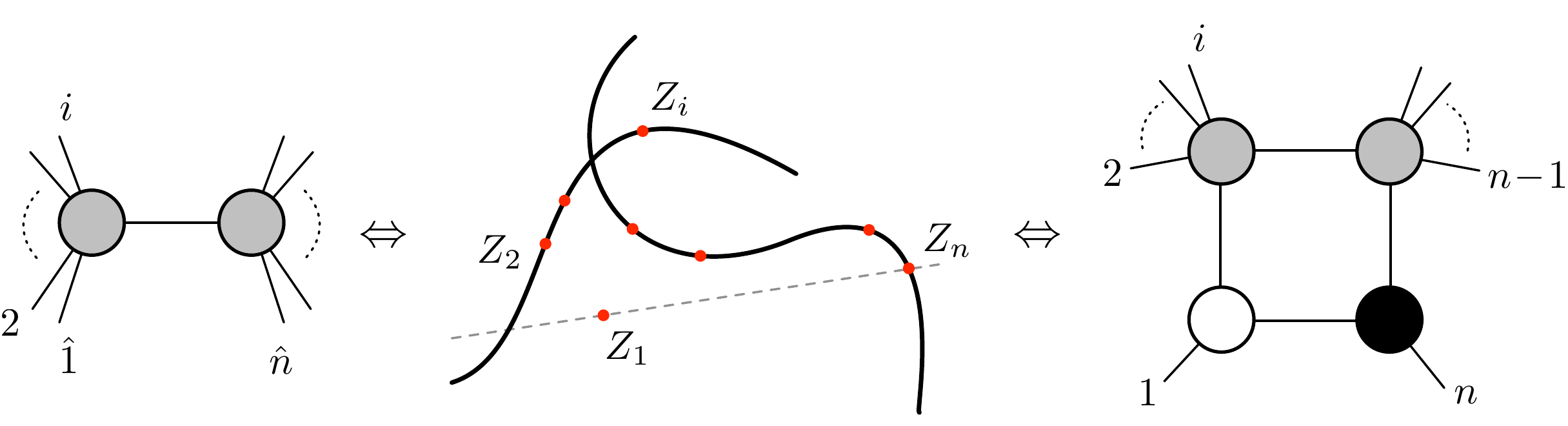}
	\caption{{\it BCFW recursion may be viewed either as a relation purely at tree level, or as a relation between tree 			amplitudes and leading singularities of (unshifted) higher loop amplitudes. These dual interpretations are clearly 			seen in the twistor geometry. }}
	\label{fig:channel}
}

Although we have focussed on the $g=0$ twistor-string in this paper, an equally valid interpretation of BCFW recursion is as a relation between tree amplitudes and certain combinations of leading singularities of loop amplitudes -- indeed, this is the way the relations were originally discovered~\cite{Britto:2004ap}. From this perspective, the external states are unshifted in the loop amplitude. A glance at figure~\ref{fig:channel} shows how the twistor geometry knows about this alternative interpretation. While the $g=0$ twistor-string worldsheet wraps only two of the curves in the picture, the first vertex operator being pulled away from $Z_1$ by the shift, we can imagine a $g=1$ twistor-string wrapping all three curves ({\it i.e.}, also wrapping the line $[Z_1\wedge Z_n]$) and thus meeting all of the external twistors. In comparison with the degree $d$, genus 0 twistor-string, the resulting configuration has degree $d+1$ and genus 1, so the Riemann-Roch theorem~\eqref{RiemRoch} shows the MHV level is unchanged. Similarly, while the $g=0$ curve has a single node, the $g=1$ configuration is of codimension four in $\overline{M}_{1,n}(\PT,d+1)$ because the worldsheet has \emph{four} nodes -- three of the worldsheet components are mapped to the three curves, while a fourth undergoes a constant map to the point $Z_n$ and corresponds to the three-point $\MHVbar$ subamplitude at the bottom right of the momentum channel diagram in figure~\ref{fig:channel}. This codimension exactly corresponds to the fact that one loop leading singularities are codimension four in momentum space.

How does this picture relate to the line configurations found in~\cite{Korchemsky:2009jv,Bullimore:2009cb} that correspond to the solution of the BCFW recursion relation? To solve the BCFW relation in twistor space, one should continue to decompose the degree $d_1$ and $d_2$ curves into curves of lower degree until everything is either MHV or $\MHVbar$.  In particular, consider the homogeneous term where $d_1=0$ and $d_2=d$ at the first step of the recursion. Then one component of the $g=0$ worldsheet -- containing the first and second marked points -- is mapped to a single point in $\PT$. This term only has support when $Z_2$ and $Z_1(t)$ coincide, or in other words when $Z_1$, $Z_2$ and $Z_n$ are collinear. Since the image of the worldsheet node also coincides with $Z_2$, if we now shift the internal state also in the direction of $Z_n$, the result of the second recursive step will look just like figure~\ref{fig:channel}, but where now $Z_1$, $Z_2$ and $Z_n$ are all collinear, with the string worldsheet intersecting this line at some new point away from $Z_1$ and $Z_2$ (as well as the intersection at $Z_n$). Continuing in this way, for sufficiently large $n$ we can transfer arbitrarily many states to the line $[Z_1\wedge Z_n]$.

At NMHV level, having solved the homogeneous term the remaining curves must each have degree 1, leading to the standard picture of an NMHV tree as a sum of 3-mass box coefficients, each of which is supported on a triangle in twistor space as in figure~\ref{fig:3mb}. At higher MHV degree, the remaining curves may still have high degree, and we must continue the recursive procedure. It is clear that the resulting products of amplitudes will have support precisely when the external twistors lie in the line configurations of~\cite{Korchemsky:2009jv,Bullimore:2009cb}. These higher genus configurations are compatible with the tree-level twistor-string once we realise that many lines are due to iterated BCFW shifts and are not wrapped by the worldsheet.

%%%%%%%%%%%%%%%%%%%%%%%%%%%%%%%%%%%%%%%%%%%%%%%%%
%%%%%%%%%%%%%%%%%%%%%%%%%%%%%%%%%%%%%%%%%%%%%%%%%

\section{Conclusions}
\label{sec:conclusions}

Twistor-string theory has always been an extremely beautiful idea that apparently suffers from a catalogue of ills. Most seriously, a worldsheet CFT that gives rise exclusively to the \emph{single trace} term we have studied in this paper is still unknown. As shown in~\cite{Berkovits:2004jj,Mason:2007zv},  the theory studied in section~\ref{sec:twsamps} also contains conformal supergravity. The close relation between the moduli space of stable maps and BCFW recursion obtained here suggests that the holomorphic sector of the worldsheet CFT is correct as it stands; perhaps only the 
correct formulation of worldsheet gravity -- and its relation with the contour in moduli space -- remains to be found.

A second, related issue that so far has received much less attention than it deserves is to understand the role of space-time signature.  In momentum space, one may continue from Lorentzian to split signature by analytically continuing the (tree-level) scattering amplitudes between the two real slices, but in twistor space the fact that the external states are really described by cohomology classes ($H^1$s rather than $H^0$s) makes this issue more subtle. In particular, Lorentzian twistor space contains a real codimension one slice $\mathbb{PN}$ that plays a fundamental role in the crucial definition of positive and negative frequency states. In the twistor-string, the choice of space-time signature is supposed to determine a path integral contour. Here we have seen that the homology classes determined by the external states themselves plays an important role in BCFW recursion, effectively reducing the necessary choice of contour to one at MHV level (as in the Grassmannian). It would be interesting to see how this is compatible with a real structure on the moduli space of higher degree maps.

In many ways, however, the most off-putting aspect of addressing these important challenges has simply been the feeling that, for all its elegance, twistor-string theory is a forbiddingly complicated object, far removed from the most successful calculational techniques in momentum space. Recent progress in relating the twistor-string to the Grassmannian contour integral~\cite{Dolan:2009wf,Bullimore:2009cb,ArkaniHamed:2009dg,Nandan:2009cc,Dolan:2010xv,Bourjaily:2010kw} has gone some way towards dispelling this belief, but the calculations required are still rather involved, especially for large numbers of particles and high MHV degree. I hope that the transparency of both the derivation and interpretation of BCFW recursion given here finally proves that the twistor-string is essentially straightforward to handle using unitarity methods, with factorization into subamplitudes corresponding to degeneration of the worldsheet in a very natural way.

Finally, let me mention that from the twistor-string recursion relation has a number of clear precedents in the string theory literature. Firstly, both its derivation and final form are closely related to a recursion relation for gravitational descendants in Gromov-Witten theory obtained in~\cite{Witten:1990hr}, whose existence may be seen in part as a consequence of the relation between two-dimensional topological gravity and the KdV hierarchy. It is also tempting to speculate that, as in~\cite{Bershadsky:1993cx} for topological strings, the $g=0$ twistor-string recursion relation extends to a recursion relation for amplitudes at higher genus, again via localisation on the boundary of the moduli space at higher genus maps. Indeed, from the twistor-string viewpoint, the conjecture that an $\ell$-loop scattering amplitude can be reconstructed from its leading singularities seems to require the existence of such a generalisation. In particular, \cite{Bullimore:2009cb} argued that the correspondence between higher genus line configurations in $\PT$ and multi-loop leading singularities could perhaps be used to bootstrap a successful twistor-string at higher genus.

Another close cousin of the twistor-string BCFW recursion relation is the equivalence between the twistor-string formula and the expression of a tree amplitude in terms of MHV diagrams~\cite{Gukov:2004ei}. I suspect that this relation can be obtained by repeating the arguments of this paper without performing a BCFW shift. The choice of which component of the $\bar\delta^{3|4}$-function to `relax' presumably corresponds to a choice of CSW reference spinor.

%%%%%%%%%%%%%%%%%%%%%%%%%%%%%%%%%%%%%%%%%%%%%%%%%
%%%%%%%%%%%%%%%%%%%%%%%%%%%%%%%%%%%%%%%%%%%%%%%%%

\vspace{1cm}

{\large\bf\noindent Acknowledgements}

\medskip

\noindent It is a pleasure to thank Nima Arkani-Hamed, Mathew Bullimore, Peter Goddard, Lionel Mason, Andy Neitzke and especially Freddy Cachazo for helpful discussions. I also thank Edward Witten for pressing me to explain more clearly how the higher genus configurations found in~\cite{Korchemsky:2009jv,Bullimore:2009cb} could be compatible with the $g=0$ twistor-string.  This work is supported by the Perimeter Institute for Theoretical Physics. Research at the Perimeter Institute is supported by the Government of Canada through Industry Canada and by the Province of Ontario through the Ministry of Research $\&$ Innovation. 

%%%%%%%%%%%%%%%%%%%%%%%%%%%%%%%%%%%%%%%%%%%%%%%%%
%%%%%%%%%%%%%%%%%%%%%%%%%%%%%%%%%%%%%%%%%%%%%%%%%
\newpage

\appendix

\section{Conventions}
\label{app:conventions}

Every paper on scattering amplitudes and twistors has a different set of conventions. This one uses the following\footnote{These conventions agree with Penrose. To translate to the spinor conventions of Wess \& Bagger one must replace ${A'}\leftrightarrow \alpha$ and $A\leftrightarrow{\dot\alpha}$ as well as dualise the R-symmetry representation. Sorry.}.

A null momentum $p$ will be represented by a bi-spinor $p_{AA'} = \tilde\pi_A\pi_{A'}$, where $A,B,\ldots$ and $A',B',\ldots$ label elements of $\mathbb{S}^-$ and $\mathbb{S}^+$, the left and right spin bundles, respectively. We will sometimes use the shorthand $[12]\equiv \epsilon^{AB}\tilde\pi_{1A}\tilde\pi_{2B}$ and $\la12\ra \equiv\epsilon^{A'B'}\pi_{1A'}\pi_{2B'}$ to denote the SL$(2,\C)$-invariant inner produces on these bundles.

The $\cN=4$ supersymmetry algebra is written as
\be
	\left\{Q_{A'a},Q^{\dagger\,b}_{A}\right\}=\delta_a^{\ b}\, p_{AA'}\,.
\ee
and on-shell may be represented by the operators
\be
	Q_{A'a} = \pi_{A'}\eta_a\qquad\hbox{and}\qquad 
	Q^{\dagger\,a}_A = \tilde\pi_A\frac{\del}{\del\eta_a}\,,
\ee
where $(\tilde\pi_A,\pi_{A'},\eta_a)$ are coordinates on on-shell momentum superspace. Since $Q^\dagger$ is a helicity raising operator, in this representation the on-shell supermultiplet
$$
	\Phi(\pi,\tilde\pi,\eta) = g^+(\pi,\tilde\pi) + \eta_a\gamma^a(\pi,\tilde\pi) + \ldots 
	+ \frac{1}{4!}\epsilon^{abcd}\eta_a\eta_b\eta_c\eta_d\, g^-(\pi,\tilde\pi)
$$
has a \emph{positive} helicity gluon as its lowest component.

We describe $\cN=4$ twistor space by the four complex bosonic and four fermionic coordinates
\be
	Z^I = (Z^\alpha,\psi^a) = (\omega^A,\pi_{A'},\psi^a)\,,
\ee
defined up to an overall scale. On twistor space, the $\cN=4$ superconformal algebra acts geometrically by Lie derivation along the vector fields 
\be
	J^I_{\ J}\equiv Z^I\frac{\del}{\del Z^J}\,,
\ee
except that the Euler vector field
\be
	\Upsilon\equiv\sum_\alpha Z^\alpha\frac{\del}{\del Z^\alpha} + \sum_a \psi^a\frac{\del}{\del\psi^a}
\ee
defines the projection $\C^{4|4}\to\CP^{3|4}$. The twistor space of (an affine patch of) split-signature space-time is (the neighbourhood of a line in) $\RP^3$, whereupon $(\omega^A,\psi^a)$ are Fourier conjugate to $(\tilde\pi_A,\eta_a)$.

On complex twistor space, the Penrose transform provides an isomorphism
\be
	\left\{
		\begin{matrix}
			\hbox{analytic solutions of helicity } h\\
			\hbox{massless free field equations}
		\end{matrix}
	\right\}
	\cong
	\left\{
		\begin{matrix}
			\hbox{elements of}\\
			 H^1(\PT,\cO(-2h-2))
		\end{matrix}
	\right\}\,.
\label{Pentrans}
\ee
In particular, twistor fields of homogeneity zero correspond to \emph{negative} helicity space-time fields, so the $\cN=4$ twistor supermultiplet
$$
	a(Z,\psi) = a^-(Z) + \psi^a\Gamma_a(Z) +\cdots	+\frac{1}{4!}\epsilon_{abcd}\psi^a\psi^b\psi^c\psi^da^+(Z)
$$
has a \emph{negative} helicity gluon as its lowest component, reflecting the fact that $(\eta_a,\psi^b)$ are conjugate variables. Correspondingly, in this paper an amplitude involving two positive and arbitrarily many negative helicity gluons is called MHV, and is supported on a line in twistor space.

%%%%%%%%%%%%%%%%%%%%%%%%%%%%%%%%%%%%%%%%%%%%%%%%%
%%%%%%%%%%%%%%%%%%%%%%%%%%%%%%%%%%%%%%%%%%%%%%%%%

\bibliographystyle{JHEP}
\bibliography{BCFWstringsref}

\providecommand{\href}[2]{#2}\begingroup\raggedright\begin{thebibliography}{10}

\bibitem{Witten:2003nn}
E.~Witten, {\it {Perturbative Gauge Theory as a String Theory in Twistor
  Space}},  {\em Commun. Math. Phys.} {\bf 252} (2004) 189--258,
  [\href{http://arxiv.org/abs/hep-th/0312171}{{\tt hep-th/0312171}}].

\bibitem{Roiban:2004yf}
R.~Roiban, M.~Spradlin, and A.~Volovich, {\it {On the Tree-Level S-Matrix of
  Yang-Mills Theory}},  {\em Phys. Rev.} {\bf D70} (2004) 026009,
  [\href{http://arxiv.org/abs/hep-th/0403190}{{\tt hep-th/0403190}}].

\bibitem{Witten:2004cp}
E.~Witten, {\it {Parity Invariance for Strings in Twistor Space}},  {\em Adv.
  Theor. Math. Phys.} {\bf 8} (2004) 779--796,
  [\href{http://arxiv.org/abs/hep-th/0403199}{{\tt hep-th/0403199}}].

\bibitem{Dolan:2009wf}
L.~Dolan and P.~Goddard, {\it {Gluon Tree Amplitudes in Open Twistor String
  Theory}},  \href{http://arxiv.org/abs/0909.0499}{{\tt arXiv:0909.0499}}.

\bibitem{Bullimore:2009cb}
M.~Bullimore, L.~Mason, and D.~Skinner, {\it {Twistor-Strings, Grassmannians
  and Leading Singularities}},  {\em JHEP} {\bf 03} (2010) 070,
  [\href{http://arxiv.org/abs/0912.0539}{{\tt arXiv:0912.0539}}].

\bibitem{ArkaniHamed:2009dg}
N.~Arkani-Hamed, J.~Bourjaily, F.~Cachazo, and J.~Trnka, {\it {Unification of
  Residues and Grassmannian Dualities}},
  \href{http://arxiv.org/abs/0912.4912}{{\tt arXiv:0912.4912}}.

\bibitem{Nandan:2009cc}
D.~Nandan, A.~Volovich, and C.~Wen, {\it {A Grassmannian Etude in NMHV
  Minors}},  \href{http://arxiv.org/abs/0912.3705}{{\tt arXiv:0912.3705}}.

\bibitem{Dolan:2010xv}
L.~Dolan and P.~Goddard, {\it {General Split Helicity Gluon Tree Amplitudes in
  Open Twistor-String Theory}},  {\em JHEP} {\bf 05} (2010) 044,
  [\href{http://arxiv.org/abs/1002.4852}{{\tt arXiv:1002.4852}}].

\bibitem{Bourjaily:2010kw}
J.~Bourjaily, J.~Trnka, A.~Volovich, and C.~Wen, {\it {The Grassmannian and the
  Twistor-String: Connecting All Trees in $\cN=4$ SYM}},
  \href{http://arxiv.org/abs/1006.1899}{{\tt arXiv:1006.1899}}.

\bibitem{ArkaniHamed:2009dn}
N.~Arkani-Hamed, F.~Cachazo, C.~Cheung, and J.~Kaplan, {\it {A Duality For The
  S Matrix}},  \href{http://arxiv.org/abs/0907.5418}{{\tt arXiv:0907.5418}}.

\bibitem{Drummond:2008cr}
J.~M. Drummond and J.~M. Henn, {\it {All Tree-Level Amplitudes in $\cN=4$
  SYM}},  {\em JHEP} {\bf 04} (2009) 018,
  [\href{http://arxiv.org/abs/0808.2475}{{\tt arXiv:0808.2475}}].

\bibitem{Korchemsky:2009jv}
G.~P. Korchemsky and E.~Sokatchev, {\it {Twistor Transform of All Tree
  Amplitudes in $\cN=4$ SYM Theory}},
  \href{http://arxiv.org/abs/0907.4107}{{\tt arXiv:0907.4107}}.

\bibitem{Britto:2004ap}
R.~Britto, F.~Cachazo, and B.~Feng, {\it {New Recursion Relations for Tree
  Amplitudes of Gluons}},  {\em Nucl. Phys.} {\bf B715} (2005) 499--522,
  [\href{http://arxiv.org/abs/hep-th/0412308}{{\tt hep-th/0412308}}].

\bibitem{Britto:2005fq}
R.~Britto, F.~Cachazo, B.~Feng, and E.~Witten, {\it {Direct Proof Of Tree-Level
  Recursion Relation In Yang-Mills Theory}},  {\em Phys. Rev. Lett.} {\bf 94}
  (2005) 181602, [\href{http://arxiv.org/abs/hep-th/0501052}{{\tt
  hep-th/0501052}}].

\bibitem{Berkovits:2004hg}
N.~Berkovits, {\it {An Alternative String Theory in Twistor Space for $\cN = 4$
  Super-Yang-Mills}},  {\em Phys. Rev. Lett.} {\bf 93} (2004) 011601,
  [\href{http://arxiv.org/abs/hep-th/0402045}{{\tt hep-th/0402045}}].

\bibitem{Berkovits:2004jj}
N.~Berkovits and E.~Witten, {\it {Conformal Supergravity in Twistor-String
  Theory}},  {\em JHEP} {\bf 08} (2004) 009,
  [\href{http://arxiv.org/abs/hep-th/0406051}{{\tt hep-th/0406051}}].

\bibitem{Mason:2007zv}
L.~Mason and D.~Skinner, {\it {Heterotic Twistor-String Theory}},  {\em Nucl.
  Phys.} {\bf B795} (2008) 105--137,
  [\href{http://arxiv.org/abs/0708.2276}{{\tt arXiv:0708.2276}}].

\bibitem{Dolan:2007vv}
L.~Dolan and P.~Goddard, {\it {Tree and Loop Amplitudes in Open Twistor-String
  Theory}},  {\em JHEP} {\bf 06} (2007) 005,
  [\href{http://arxiv.org/abs/hep-th/0703054}{{\tt hep-th/0703054}}].

\bibitem{Brandhuber:2008pf}
A.~Brandhuber, P.~Heslop, and G.~Travaglini, {\it {A Note on Dual
  Superconformal Symmetry of the $\mathcal{N}=4$ Super-Yang-Mills $S$-Matrix}},
   {\em Phys. Rev.} {\bf D78} (2008) 125005,
  [\href{http://arxiv.org/abs/0807.4097}{{\tt arXiv:0807.4097}}].

\bibitem{ArkaniHamed:2008gz}
N.~Arkani-Hamed, F.~Cachazo, and J.~Kaplan, {\it {What is the Simplest Quantum
  Field Theory?}},  \href{http://arxiv.org/abs/0808.1446}{{\tt
  arXiv:0808.1446}}.

\bibitem{Mason:2009sa}
L.~Mason and D.~Skinner, {\it {Scattering Amplitudes and BCFW Recursion in
  Twistor Space}},  \href{http://arxiv.org/abs/0903.2083}{{\tt
  arXiv:0903.2083}}.

\bibitem{Bershadsky:1993cx}
M.~Bershadsky, S.~Cecotti, H.~Ooguri, and C.~Vafa, {\it {Kodaira-Spencer theory
  of gravity and exact results for quantum string amplitudes}},  {\em Commun.
  Math. Phys.} {\bf 165} (1994) 311--428,
  [\href{http://arxiv.org/abs/hep-th/9309140}{{\tt hep-th/9309140}}].

\bibitem{Kapranov:modulicurves}
M.~Kapranov, {\it {Veronese curves and the Grothendieck-Knudsen moduli space
  $\overline{M}_{0,n}$}},  {\em J.~Alg.~Geom.} {\bf 2} (1993) 239--262.

\bibitem{Witten:1990hr}
E.~Witten, {\it {Two-dimensional Gravity and Intersection Theory on Moduli
  Space}},  {\em Surveys in Diff. Geom.} {\bf 1} (1991) 243--310.

\bibitem{Vergu:2006np}
C.~Vergu, {\it {On the Factorisation of the Connected Prescription for
  Yang-Mills Amplitudes}},  {\em Phys. Rev.} {\bf D75} (2007) 025028,
  [\href{http://arxiv.org/abs/hep-th/0612250}{{\tt hep-th/0612250}}].

\bibitem{Gukov:2004ei}
S.~Gukov, L.~Motl, and A.~Neitzke, {\it {Equivalence of Twistor Prescriptions
  for Super Yang-Mills}},  {\em Adv. Theor. Math. Phys.} {\bf 11} (2007)
  199--231, [\href{http://arxiv.org/abs/hep-th/0404085}{{\tt hep-th/0404085}}].

\end{thebibliography}\endgroup

\end{document}